# Accurate THz Measurements of Permittivity and Permeability of BiFeO$_3$ Thin Films

Gian Paolo Papari[1,2,*], Zahra Mazaheri[1], Francesca Lo Presti[3], Graziella Malandrino[3], and Antonello Andreone[1,2]

[1]Dipartimento di Fisica, Università di Napoli "Federico II," via Cinthia, I-80126 Napoli, Italy
[2]CNR-SPIN, UOS Napoli, via Cinthia, I-80126 Napoli, Italy
[3]Dipartimento di Scienze Chimiche, Università di Catania, and INSTM UdR Catania, Viale A. Doria 6, I-95125 Catania, Italy;

*gianpaolo.papari@unina.it

**Abstract.** The electrodynamic properties of BiFeO$_3$ films in the THz region are investigated via time domain spectroscopy. Combining the use of transmission ($\tilde{T}$) and reflection ($\tilde{R}$) measurements under normal incidence, the refractive index and impedance of the samples under test are evaluated using a retrieval method. From $\tilde{T}$ ($\tilde{R}$) data two complex functions describing the refractive index $\tilde{n}_T$ ($\tilde{n}_R$) and impedance $\tilde{z}_T$ ($\tilde{z}_R$) are extracted from the independent minimization of the error functions given by the difference between the theoretical model and measurements. Knowledge of the pairs ($\tilde{n}_T$, $\tilde{n}_R$) and ($\tilde{z}_T$, $\tilde{z}_R$) enables to calculate with a high accuracy both complex permittivity $\tilde{\varepsilon}$ and permeability $\tilde{\mu}$ of the sample. Signatures of magnetoelectric effects and phononic resonances are observed in the permittivity and permeability functions and discussed in detail.

## 1. Introduction

BiFeO$_3$ (BFO) is a multiferroic material [1] that for several years has attracted the interest of researchers, owed to the multiple applications where it can be employed. The combination of ferroelectric and ferromagnetic properties makes this material a promising candidate for the development of novel technologies with high performance. Energy storage [2] [3] [4], spintronics [5] [6] [7], photovoltaics [8] [9] [10] [11] [12], coherent radiation emitters [13] [14] are among the fields where BFO can find a remarkable role. Recently, different spectroscopic investigations have been performed to study the electrical and magnetic response of the material. To mention a few, soft mode [15] microwave [16], and Raman spectroscopy [17] [18] [19]. X-ray spectroscopy has been also largely used to correlate the BFO ferroelectric and magnetic properties to the structure of orbitals [20] [21] [22] [23]. In this framework, Terahertz (THz) spectroscopy assumes a special role in the investigations of BFO excitations since it directly probes specific energy states which can be exploited for the next generation of high frequency switchable devices. BFO in fact presents different resonances in the THz spectrum. Below 2 THz, magnetic (magnons) and dipole resonances may be observed [24] [25] [26], whereas above 2 THz robust phonon-mediated peaks are present [17] [23] [27]. The magnetoelectric coupling in the microwave band has been thoroughly studied, making possible the extraction of both the electrical (complex permittivity $\tilde{\varepsilon}$) and magnetic (complex permeability $\tilde{\mu}$) response [16]. In the THz band, the phenomenon has been weakly addressed since a reliable procedure to extract $\tilde{\varepsilon}$ and $\tilde{\mu}$ independently is lacking [25] [28]. Indeed, the antiferromagnetic behavior of BFO has been mostly measured through magnetostatic measurements [1] [29] yielding $\tilde{\mu} \sim 1$. Since $\tilde{n} = \sqrt{\tilde{\mu}\tilde{\varepsilon}}$, where $\tilde{n} = n + ik$ is the complex refractive index and $\tilde{\varepsilon} = \varepsilon_r + i\varepsilon_i$ is the complex permittivity, the result $\tilde{\mu} \sim 1$ has been extended also to THz frequencies [24] [25], leading to the standard equation $\tilde{n}^2 = \tilde{\varepsilon}$. Nevertheless, at present a full comprehension of the magnetic response of BFO at very high frequencies is still missing.

To retrieve both the permeability and permittivity functions of BFO thin films in the THz range, we have developed an accurate methodology to extract $\tilde{\varepsilon}$ and $\tilde{\mu}$ from complex transmission ($\tilde{T}$) and reflection ($\tilde{R}$) measurements via time domain spectroscopy (TDS). Using a combined total variation technique (CTVT) [30], we can independently extract $\tilde{n}$ and the complex impedance $\tilde{z}$ of the sample under test in a range delimited by the pairs ($\tilde{n}_T$, $\tilde{n}_R$) and ($\tilde{z}_T$, $\tilde{z}_R$), respectively acquired from the measured $\tilde{T}$ and $\tilde{R}$ functions, providing a confidence interval in which a precise estimation of $\tilde{\varepsilon} = \tilde{n}\,\tilde{z}$ and $\tilde{\mu} = \tilde{n}/\tilde{z}$ can be obtained.

TDS is an experimental technique capable of measuring the complex quantities $\tilde{T}$ and $\tilde{R}$ of a bulk sample in the terahertz band, provided that its thickness fulfils the requirements for a suitable detection. In particular, thickness should not exceed the material skin depth [31] to avoid the transmitted signal be falling below the noise $\tilde{T} \sim 0$. At the same time, it must not be too small as well to prevent the case $\tilde{R} \sim 0$. The advantage of TDS spectroscopy is its coherent signal detection of the signal, which allows for the acquisition of both the modulus and phase of the EM signal in transmission and reflection mode. By means of these latter quantities, $\tilde{n}$ and $\tilde{z}$ [32], the permittivity and permeability can be extracted. In literature, various approaches can be found to retrieve the electrodynamic parameters of a material in the THz band by using time domain experiments [30] [33] [34] [35] [36] [37] [38]. However, they either do not provide simultaneous measurements of the transmission and reflection functions or use simulated $\tilde{T}$ and $\tilde{R}$ functions [39]. It is worth noting that the retrieval of $\tilde{\varepsilon}$ and $\tilde{\mu}$ operated in the microwave band by means of the scattering parameters $\tilde{S}_{11}$ and $\tilde{S}_{12}$ [40] [41] [42] [43] is generally not compatible with free space TDS-THz experiments since it implies a completely different experimental approach, usually based on the use of waveguides to test the response of the sample under test.

To the best of our knowledge, an accurate method to retrieve $\tilde{\varepsilon}$ and $\tilde{\mu}$ of an unknown material in the THz band is introduced by Nemec *et al*. in [32], where thick samples are investigated via TDS to first get $\tilde{z}$ from $\tilde{R}$ and then $\tilde{n}$ from $\tilde{T}$. Inspired by this work, we have developed a similar approach capable to achieve $\tilde{z}$ and $\tilde{n}$ curves as a function of frequency using modelled transmission and reflection complex functions. The method predicts the experimental data with excellent accuracy. The retrieval process is applied to thin films of BFO grown on SiO$_2$ substrates. The material electrodynamics is studied in the band 0.4-2.5 THz and its behavior as a function of frequency discussed, shedding a light from one side on the electric and magnetic coupling with the EM radiation, and on the other on the magnetoelectric phenomena showing up in correspondence of the phononic resonance at around 2 THz. The paper is divided into several parts: following this brief introduction, in section 2 the theoretical background underneath the retrieval process for thick and thin samples is discussed. In section 3 the measurement setup is illustrated, whereas in section 4 the fabrication and the morphological/crystallographic characterization of the BFO samples is reported. Sections 5 and 6 are devoted to the description of the permittivity and permeability of the substrate and film respectively. In section 7 a discussion on the electric and magnetic properties of the BFO films is presented and in section 8 a number of conclusions are drawn.

## 2. Theoretical background

### 2.1 Retrieving $\tilde{z}$ and $\tilde{n}$ for a bare substrate

The transmission and reflection functions are modelled through the use of Fresnel coefficients and propagation factors [44]. Assuming a linearly polarized EM planar wave with angular frequency $\omega = 2\pi f$ propagates from medium $\alpha$ across medium $\beta$, the two Fresnel coefficients accounting for the amount of the transmitted and reflected wave are $\tilde{t}_{\alpha,\beta} = 2\frac{\tilde{z}_\beta}{\tilde{z}_\alpha+\tilde{z}_\beta}$ and $\tilde{r}_{\alpha,\beta} = \frac{\tilde{z}_\alpha-\tilde{z}_\beta}{\tilde{z}_\alpha+\tilde{z}_\beta}$, where $\tilde{z}_{\alpha,\beta}$ is the complex impedance of the two media normalized to the vacuum impedance $\tilde{z}_0 = 377\ \Omega$. Passing through medium $\beta$, the EM wave experiences a phase delay and attenuation described by the complex propagation factor $\tilde{P}_{\beta,d} = \exp i\omega \tilde{n}_\beta d/c$, where $\tilde{n}_\beta = n_\beta + ik_\beta$ is the complex refractive index of the material, assumed homogeneous, having thickness $d$, and $c$ is the speed of light in vacuum.

According to the procedure discussed in [45], the signal transmitted in air passing across a single homogeneous slab of thickness $d$ can be obtained employing the concept of fundamental signal $\tilde{E}_{T,F} = \tilde{E}_{T,in}\tilde{t}_{a,s}\tilde{P}_{s,d}\tilde{t}_{s,a}$. This term depends only on the impinging signal $\tilde{E}_{T,in}$, the Fresnel coefficients and the propagation factor, and neglects multiple (internal) reflections. Subscripts $a$ and $s$ refer to air ($\tilde{n}_a = 1$) and slab respectively. The signal transmitted in air (no sample) represents the reference signal $\tilde{E}_{T,ref} = \tilde{E}_{T,in}\tilde{P}_{a,d}$.

The transmitted signal is sketched in Fig.1(a). Since the total transmitted electric field is $\tilde{E}_T = \tilde{E}_{T,F}\widetilde{FP}_s$, where $\widetilde{FP}_s = 1/(1 - \tilde{r}_{s,a}^2 \tilde{P}_{s,d}^2)$ is the Fabry-Perot (FP) term, the transmission function of the slab is given by:

$$\tilde{T}_s = \frac{\tilde{E}_T}{\tilde{E}_{T,ref}} = \frac{\tilde{t}_{a,s}\tilde{t}_{s,a}\ e^{i\omega(\tilde{n}_s-1)d/c}}{1-\tilde{r}_{s,a}^2 e^{i\omega \tilde{n}_s 2d/c}} \tag{1}$$

The scheme for the reflected signal is reported in Fig. 1(b). The reflection function is composed of two terms, given by the interaction with the first and second interface. In the latter case it has to consider the presence of multiple reflections. The total reflected signal is $\tilde{E}_R = \tilde{E}_{R,in}\tilde{r}_{a,s} + \tilde{E}_{R,in}\tilde{t}_{a,s}P_{s,d}^2\tilde{r}_{s,a}\tilde{t}_{s,a}\widetilde{FP}_s = \tilde{E}_{R,in}\tilde{r}_{a,s} + \tilde{E}_{R,in}\tilde{E}_{R,F}\widetilde{FP}_s$, where $\tilde{E}_{R,F}$ is the fundamental reflected signal. The reflected reference signal is acquired employing a metallic (gold) mirror, assumed being ideally conducting, and is given by $\tilde{E}_{R,ref} = \tilde{E}_{R,in}\tilde{P}_{a,L}^2$. Since the distance mirror-detector is usually different from the distance sample-detector, the reference phase contains the delay $\exp \pm i\omega\delta t$, where $\delta t = 2L/c$ is proportional to the optical path $L$ between the sample and mirror surface. The sign of the phase is positive/negative depending on the mirror distance being farther/closer than the sample first interface. Therefore, the full expression of the reflection function of the slab is the following:

$$\tilde{R}_s = \frac{\tilde{E}_R}{\tilde{E}_{R,ref}} = \left(\tilde{r}_{a,s} + \frac{\tilde{t}_{a,s}\ e^{\frac{i\omega 2\tilde{n}_s d}{c}}\tilde{r}_{s,a}\tilde{t}_{s,a}}{1-\tilde{r}_{s,a}^2 e^{\frac{i\omega \tilde{n}_s 2d}{c}}}\right)/e^{\pm i\omega\delta t}. \tag{2}$$

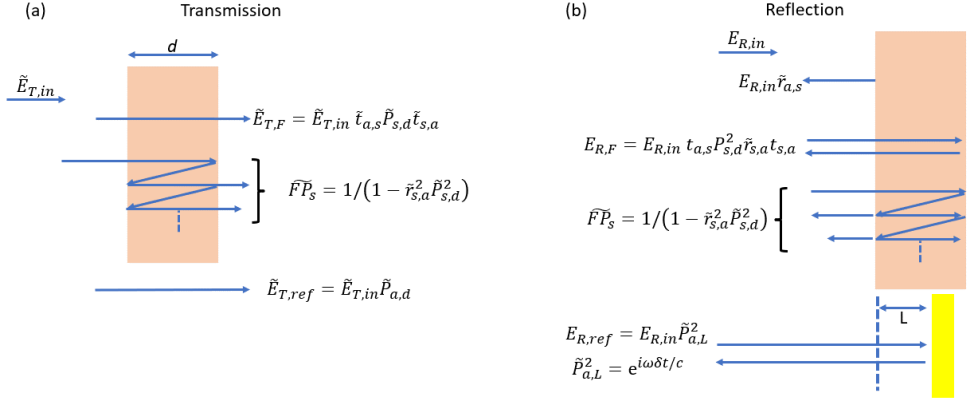

*Figure 1: Sketch of (a) the transmitted and (b) the reflected EM signals normally impinging on a single homogeneous slab of thickness d, together with the corresponding theoretical expressions, term by term. The yellow slab in (b) represents an ideally conducting (gold) mirror.*

Therefore, given the full theoretical expressions of $\tilde{T}_s$ and $\tilde{R}_s$, it is clear that the target parameters $\tilde{z}_s$ and $\tilde{n}_s$ cannot be easily extracted through a simple manipulation of eqs. (1) and (2). Core of our strategy is to start from approximate expressions of transmission and reflection functions, which we name $\tilde{T}_{s0}$ and $\tilde{R}_{s0}$, to enable the analytical extraction of the zero-order parameters $\tilde{z}_{s0}$ and $\tilde{n}_{s0}$. The approximation consists in neglecting the FP terms so turning eqs. (1) and (2) into

$$\tilde{R}_{s0} = \tilde{r}_{a,s}/\tilde{P}_{a,L}^2 = \frac{\tilde{z}_a - \tilde{z}_{s0}}{\tilde{z}_a + \tilde{z}_{s0}} e^{-i\omega\delta t} \qquad (3)$$

$$\tilde{T}_{s0} = \frac{\tilde{t}_{a,s}\tilde{P}_{s,d}\tilde{t}_{s,a}}{\tilde{P}_{a,d}} = 4\frac{\tilde{z}_a \tilde{z}_{s0}}{(\tilde{z}_a + \tilde{z}_{s0})^2} e^{i\omega(\tilde{n}_{s0}-1)d/c} \qquad (4)$$

where $\tilde{z}_a = 1$.

From eq. (3) the slab impedance results

$$\tilde{z}_{s0} = \tilde{z}_a \frac{1 - \tilde{r}_{s0}}{1 + \tilde{r}_{s0}} \qquad (5)$$

where $\tilde{r}_{s0} = \tilde{R}^{(e)}_{s0} e^{i\omega\delta t}$, $\tilde{R}^{(e)}_{s0}$ being the experimental reflection function.

The achieved $\tilde{z}_{s0}$ is then used to extract the zero-order refractive index:

$$\tilde{n}_{s0} - 1 = \frac{c}{\omega d}\left[\left(\phi_{T^{(e)}s0} - \phi_\alpha + 2\pi s\right) + i\left(\ln|\tilde{\alpha}| - \ln|\tilde{T}^{(e)}_{s0}|\right)\right] \qquad (6)$$

where $\tilde{\alpha} = |\tilde{\alpha}|e^{i\phi_\alpha} = \tilde{t}_{a,s}\tilde{t}_{s,a}$, $\phi_{T^{(e)}s0}$ is the phase of the experimental transmission function $\tilde{T}^{(e)}_{s0}$, and $s = 0, \pm 1, \pm 2 \ldots$ is the refractive index branch. The latter can be identified by comparing the order of magnitude of $n_{s0}$ with the experimental value retrievable through the delay $\Delta t$ of the time dependent signal passing through the sample with respect to the reference signal. $\Delta t$ is related to the approximated refractive index through the expression $n_{s0} - 1 = c\Delta t/d$.

The experimental $\tilde{T}^{(e)}_{s0}$ and $\tilde{R}^{(e)}_{s0}$ curves used to extract the zero-order impedance and refractive index in eqs. (3) and (4) are correspondingly processed to remove the FP contribution, using an appropriate adjacent averaging procedure as described in detail in Appendix I.

Once $\tilde{z}_{s0}$ and $\tilde{n}_{s0}$ are obtained, they are used as starting functions to apply the CTVT process.

CTVT is based on the use of the full expression of the theoretical transmission and reflection functions as reported in eqs. (1) and (2). It allows to extract the best $\tilde{n}_s$ and $\tilde{z}_s$ by the minimization of two independent error functions based on reflection and transmission measurements:

$$Err(\tilde{R}_s) = \left(Re\{\tilde{R}_s^{(e)}\} - Re\{\tilde{R}_s\}\right)^2 + \left(Im\{\tilde{R}_s^{(e)}\} - Im\{\tilde{R}_s\}\right)^2 \tag{7}$$

$$Err(\tilde{T}_s) = \left(Re\{\tilde{T}_s^{(e)}\} - Re\{\tilde{T}_s\}\right)^2 + \left(Im\{\tilde{T}_s^{(e)}\} - Im\{\tilde{T}_s\}\right)^2. \tag{8}$$

The variational approach is described using the flowchart shown in Fig. 2, and it implies the independent minimization of $Err(\tilde{R}_s)$ and $Err(\tilde{T}_s)$ through a procedure that enables to retrieve the best pairs $(\tilde{n}_{s,R}, \tilde{z}_{s,R})$ and $(\tilde{n}_{s,T}, \tilde{z}_{s,T})$ respectively.

The method follows different steps. It starts from $\tilde{R}_{s0}$, which is the function with the larger approximation since it loses an addend in the removal of the FP term. We first improve the accuracy of the impedance $\tilde{z}_{s0}$ (see eq. (5)) to achieve $\tilde{z}_{s1,R}$ and $\tilde{z}_{s1,T}$ by imposing the minimization of the error functions in eqs. (7) and (8), while keeping the value of the refractive index fixed at $\tilde{n}_s = \tilde{n}_{s0}$. Afterwards, we get the terms $\tilde{n}_{s1,R}$ and $\tilde{n}_{s1,T}$ by minimizing the corresponding error functions while keeping the value of impedance fixed at $\tilde{z}_{s0}$. Following a seminal paper by Duvillaret *et al.* [30], in each minimization process the parameter subject to the improvement of accuracy is varied within an appropriate range, say $\tilde{z}_{s1} = \tilde{z}_{s0} \pm \Delta \tilde{z}_s$ to allow – for each frequency– the lowest value in eqs. (7) and (8). In the subsequent step $\tilde{z}_{s2,R}$ and $\tilde{z}_{s2,T}$ are achieved by fixing the refractive index to $\tilde{n}_{s1,R}$ and $\tilde{n}_{s1,T}$ in the respective error functions that must be minimized. The process holds until the convergence of the impedance and refractive index is fulfilled for both error functions.

To estimate how reliable the retrieved parameters are, we introduce, in the iterative procedure two "matching" factors $F_m$ for the transmission and reflection functions expressing the convergence between the model and the experimental curves as a function of the couple $\tilde{z}_j, \tilde{n}_j$, where $j$ is the indexed step. For instance, in the case of reflection we evaluate:

$$F_{m,R}(\tilde{z}_j, \tilde{n}_j) = \sum_i \left(\left|R_s^2(\tilde{z}_j, \tilde{n}_j) - R_s^{(e)^2}\right|/R_s^2(\tilde{z}_j, \tilde{n}_j)\right) \tag{9}$$

$$F_{m,\arg(R)}(\tilde{z}_j, \tilde{n}_j) = \sum_i \left(\left|\arg\left(R_s(\tilde{z}_j, \tilde{n}_j)\right) - \arg(R_s^{(e)})^2\right|/R_s^2(\tilde{z}_j, \tilde{n}_j)\right) \tag{10}$$

for the modulus and argument respectively of the complex reflection function. The same formulas apply in the case of the transmission function. In the sum the index *i* runs over the curve points. In practice, after multiple runs the pair $(\tilde{z}_R, \tilde{n}_R)$ fulfilling the inequality $F_{m,R}(\tilde{z}_R, \tilde{n}_R) \sim F_{m,\arg(R)}(\tilde{z}_R, \tilde{n}_R) < 1\%$ is chosen. The same procedure applies in transmission. Once accurate expressions of impedance and refractive index are achieved, $\tilde{\varepsilon}$ and $\tilde{\mu}$ will be defined within a confidence interval bounded by the curves $\tilde{\varepsilon}_T = \tilde{n}_T/\tilde{z}_T$ and $\tilde{\varepsilon}_R = \tilde{n}_R/\tilde{z}_R$ for the permittivity, $\tilde{\mu}_T = \tilde{n}_T \tilde{z}_T$ and $\tilde{\mu}_R = \tilde{n}_R \tilde{z}_R$ for the permeability.

The extracted refractive index and impedance are obtained by setting the best estimation on the sample thickness $d = d_0$. The uncertainty in the sample thickness ($\pm \delta d$) primarily determines the error $\Delta \tilde{z}$ in the impedance and $\Delta\tilde{n}$ in the refractive index, so that $\tilde{z} = \tilde{z}(d_0) \pm \Delta\tilde{z}(d_0 \pm \delta d)$ and $\tilde{n} = \tilde{n}(d_0) \pm \Delta\tilde{n}(d_0 \pm \delta d)$. The relative error of both real and imaginary parts of permittivity and permeability is then calculated through the root mean square of the relative errors in $\tilde{z}$ and $\tilde{n}$ since $\frac{\delta\tilde{\varepsilon}_{T,R}}{\tilde{\varepsilon}_{T,R}} = \frac{\delta\tilde{\mu}_{T,R}}{\tilde{\mu}_{T,R}} = \sqrt{\left(\frac{\delta\tilde{z}_{T,R}}{\tilde{z}_{T,R}}\right)^2 + \left(\frac{\delta\tilde{n}_{T,R}}{\tilde{n}_{T,R}}\right)^2}$.

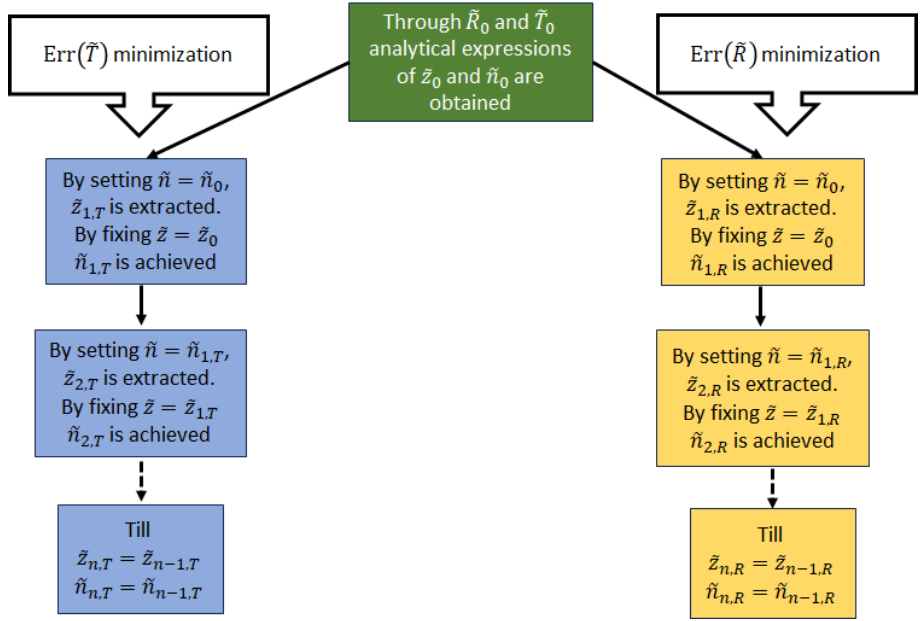

Figure 2: Flowchart of the Combined Total Variation Technique (CTVT) process. $\tilde{R}_0$ and $\tilde{T}_0$ represent the reflection and the transmission functions without the FP contribution, whereas $\tilde{T}$ and $\tilde{R}$ are the full expressions given by eqs. (1) and (2) respectively.

## 2.2 Film over the substrate

The retrieved parameters $\tilde{z}_s$ and $\tilde{n}_s$ of the substrate are then used to extract the impedance $\tilde{z}_f$ and the refractive index $\tilde{n}_f$ of the film (having thickness $t$) grown on it. Following previous approaches [34] [46], transmission and reflection mechanisms are accounted for introducing in the model the FP term for the film only and neglecting its contribution in the substrate. Thus, to remove the FP term originated from the pulse oscillations in the substrate, the experimental curves $\tilde{T}_f^{(e)}$ and $\tilde{R}_f^{(e)}$ are properly averaged to remove the sinusoidal oscillation from the modulus and phase, as described in Appendix I.

In the sketch of Fig. 3 the fundamental signal $\tilde{E}_{T,F}$, the Fabry-Perot contribution from the film $\widetilde{FP}_f$, and the reference signal $\tilde{E}_{T,ref}$ (passing through the substrate only) are described.

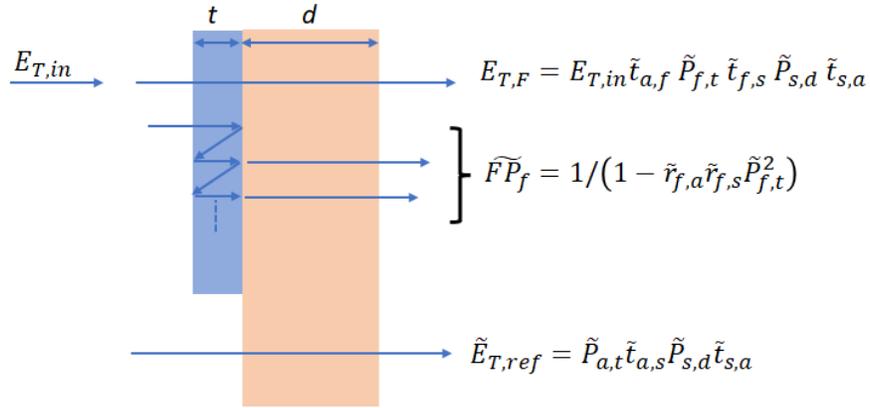

Figure 3: Sketch of the EM signal normally impinging and passing through a thin film deposited over a thick substrate, or through the substrate only (single homogeneous slab of thickness d), together with the corresponding theoretical expressions, term by term.

Since $\tilde{E}_{T,F} = \tilde{E}_{T,in}\tilde{t}_{a,f}\tilde{P}_{f,t}\tilde{t}_{f,s}\tilde{P}_{s,d}\tilde{t}_{s,a}$, and the FP term in the film reads $\widetilde{FP}_f = 1/(1-\tilde{r}_{f,a}\tilde{r}_{f,s}\tilde{P}_{f,t}^2)$, we get the following transmission function for the thin film:

$$\tilde{T}_f = \frac{\tilde{t}_{a,f}\tilde{t}_{f,s}\,e^{i\omega(\tilde{n}_f-1)t/c}}{\tilde{t}_{a,s}}\left(\frac{1}{1-\tilde{r}_{f,a}\tilde{r}_{f,s}e^{\frac{i\omega\tilde{n}_f 2t}{c}}}\right) \quad (11)$$

consistent with formulas found in other works ( [35] and reference therein), provided that the Fresnel coefficients are expressed in terms of the air, sample and substrate impedance.

Similarly for the reflection, in the sketch of Fig. 4 the fundamental signal $\tilde{E}_{R,F}$, the Fabry-Perot contribution from the film $\widetilde{FP}_f$, and the reference signal $\tilde{E}_{R,ref}$ (reflected by the substrate only) are described.

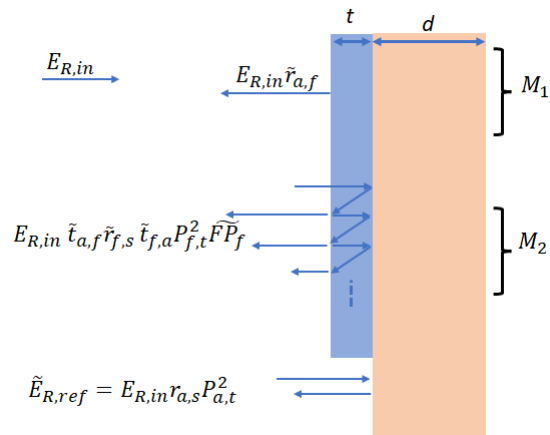

Figure 4: Sketch of the EM signal normally impinging and reflected by a thin film deposited over a thick substrate, or by the substrate only, together with the corresponding theoretical expressions, term by term.

.

According to Fig.4, the reflection function must include the two interfaces air-film and film-substrate, and the corresponding FP contribution in the film, yielding

$$\tilde{R}_f = \frac{1}{\tilde{r}_{a,s}e^{i\omega 2t/c}} \left[ \tilde{r}_{a,f} + \tilde{t}_{a,f} e^{\frac{i\omega \tilde{n}_f 2t}{c}} \tilde{r}_{f,s} \tilde{t}_{f,a} \frac{1}{1-\tilde{r}_{f,a}\tilde{r}_{f,s}e^{\frac{i\omega \tilde{n}_f 2t}{c}}} \right], \qquad (12)$$

where we replace the reference signal $\tilde{E}_{R,ref} = \tilde{E}_{in}\tilde{r}_{a,s}e^{i\omega \delta t_f}$, being $\delta t_f = 2t/c$.

In a similar fashion to the approach reported for the case of a bare substrate, the retrieval procedure of the electrodynamic parameters for a thin film starts from neglecting the FP terms and achieving an approximate analytical expression of transmission and reflection through the functions

$$\tilde{R}_{f0} = \frac{\tilde{r}_{a,f}}{\tilde{E}_{R0,ref}} = \frac{\tilde{r}_{a,f}}{\tilde{r}_{a,s}} e^{-i\omega \delta t_f} = \frac{\tilde{z}_a - \tilde{z}_{f0}}{\tilde{z}_a + \tilde{z}_{f0}} \frac{\tilde{z}_a + \tilde{z}_s}{\tilde{z}_a - \tilde{z}_s} e^{-i\omega \delta t_f} \qquad (13)$$

$$\tilde{T}_{f0} = \frac{\tilde{E}_{T,F}}{\tilde{E}_{T0,ref}} = \frac{\tilde{t}_{a,f}\tilde{t}_{f,s} e^{i\omega(\tilde{n}_f - 1)t/c}}{\tilde{t}_{a,s}} 2 \frac{\tilde{z}_{f0}}{\tilde{z}_{f0}+\tilde{z}_a} \frac{\tilde{z}_a+\tilde{z}_s}{\tilde{z}_{f0}+\tilde{z}_s} e^{i\omega(\tilde{n}_f - 1)t/c}. \qquad (14)$$

From eq. (13) we yield the impedance $\tilde{z}_{f0}$:

$$\tilde{z}_{f0} = \tilde{z}_a \left( \frac{1+\tilde{r}_f}{1-\tilde{r}_f} \right) \qquad (15)$$

where $\tilde{r}_f = \tilde{R}_f^{(e)} \tilde{r}_{a,s} e^{i\omega \delta t_f}$.

Substituting $\tilde{z}_{f0}$ in eq. (14) we can obtain the analytical expression of the refractive index:

$$\tilde{n}_{f0} - 1 = \frac{c}{\omega t} \left[ \left( \phi_{T_{f0}^{(e)}} - \phi_\alpha \right) + i \left( \ln|\tilde{\alpha}| - \ln T_{f0}^{(e)} \right) \right] \qquad (16)$$

where $\tilde{\alpha} = \tilde{t}_{a,f}\tilde{t}_{f,s}/\tilde{t}_{a,s}$.

The subsequent step is to apply the recursive procedure of the CTVT to minimize the error functions $Err(\tilde{T}_f)$, $Err(\tilde{R}_f)$ which are given by the theoretical curves described by eqs. (7) and (8).

Since we are dealing with a thick substrate, in eq. (16) we do not introduce the branch index $s$ as in the case described by eq. (6), because the $t \ll d$ usually holds and renders $s = 0$. This assumption can be qualitatively understood by evaluating the average phase shift $<\phi_n> = \frac{\omega(<n_f>-1)t}{c}$ as a function of the average refractive index $<n_f> \sim 1 + \frac{c\Delta t_f}{t} \sim 7$, where $\Delta t_f \approx 80\ fs$ is the delay measured from the time dependent signals passing through the bare substrate and the sample with film. It is easy to verify that in the working frequency band $[0.4, 2.5]$THz, the inequality $\frac{<\phi_n>}{2\pi} < 1$ is fully satisfied.

Analogously to the case with the substrate only, $\tilde{n}_{f0}$ and $\tilde{z}_{f0}$ are then used to run the CTVT by employing appropriate intervals of $\Delta \tilde{n}_f$ and $\Delta \tilde{z}_f$ to minimize $Err(\tilde{T}_f)$, $Err(\tilde{R}_f)$.

## 3. Measurement Setup

THz time domain spectroscopy measurements are routinely performed employing a commercial system (TERA K15 by Menlo Systems[®]). THz waves are generated and detected by using photo-antennas capable of performing coherent spectroscopy within the frequency range 0.1-4.0 THz. Time dependent signals are acquired for a duration of about 200 ps, providing in such a way a frequency resolution of 5 GHz. Measurements are conducted in a purging box filled with $N_2$ gas to wash out $H_2O$ absorptions.

In Fig. 5(a) the transmission set-up is shown, where the generated THz beam is first collimated and then focused, at the last impinging orthogonally on the sample surface. In Fig. 5(b) the reflection set-up is displayed. Here measurements are performed employing a 45° tilted Si beam splitter (BS) that allows to record the signal normally reflected from the sample.

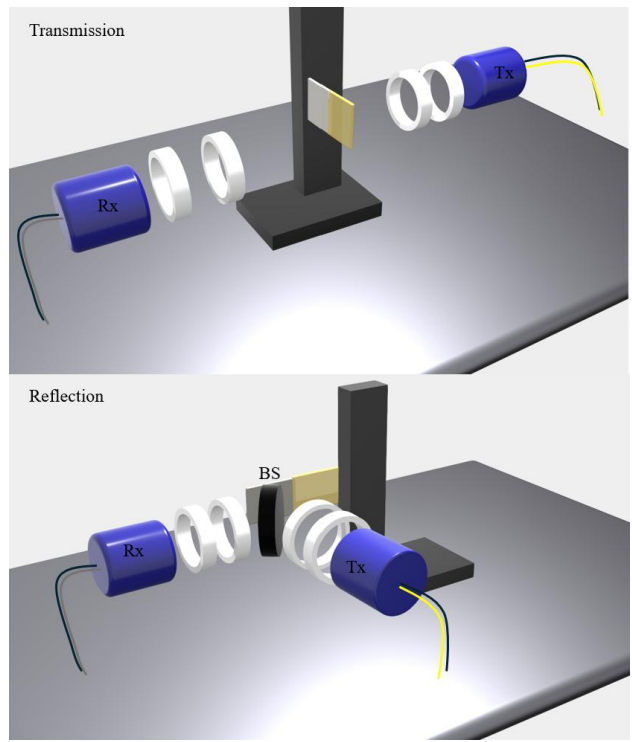

*Figure 5: Sketch of the measurement optical setup in (a) normal transmission and (b) normal reflection (with a Si beam splitter BS) configuration.*

To record time dependent signals complying with the phase shifts reported in the model discussed above, each sample consists of a BFO thin film deposited on half of the area of a 1 x 2 $cm^2$ $SiO_2$ substrate. The other half is the bare substrate only and is used to acquire the reference signal in both transmission and reflection mode. A stage shifter allows to acquire all needed signals for the transmission and reflection measurements in the same experimental session. This procedure is indispensable to minimize misalignments and related issues, as properly discussed in Sections 5 and 6.

## 4. Fabrication and morphological characterization of BFO films

BFO films have been deposited through a Molecular Organic Chemical Vapor Deposition (MOCVD) approach on the half side of a 1 x 2 cm² $SiO_2$ substrate by following the procedure reported in [47]. The precursor combination, comprising $Bi(phenyl)_3$ and $Fe(tmhd)_3$, was heated to 120 °C in an alumina crucible. The deposition was carried out in a hot-wall MOCVD reactor under reduced pressure, using argon (150 sccm) as carrier gas and oxygen as reactive gas. The films were synthesized at a temperature of 750 °C for a duration time of 60 minutes. The structure was determined through X-ray diffraction (XRD) measurements in grazing incidence (0.5°) mode, using a Rigaku Smartlab diffractometer with a Cu $K_\alpha$ rotating anode operating at 45 kV and 200 mA. The XRD pattern (Fig. 6(a)) has shown clear, well-defined peaks that closely match the polycrystalline $BiFeO_3$ structure (PDF Card No. 20-0169). The distinctive peaks at 22.48°, 31.78°, 39.48°, 45.84°, 51.40°, and 57.16° are assigned to reflections indexed as 100, 110, 111, 200, 210, and 211, respectively, for a rhombohedral structure that belongs to the R3c space group. The peak positions closely align with the ICDD reference data; however, the peak intensities differ from the ICDD reference data. Interestingly, the 100 reflection peak at 22.48° is the strongest one, indicating a preferential orientation along this direction. The morphologies of the films have been examined through Field Emission Scanning Electron Microscopy (FE-SEM), using a ZEISS SUPRA 55 VP microscope. The films have good adhesion and homogeneous features across the substrate. As observed in Fig. 6(b), the BFO films showed well-coalesced grains in the order of hundreds of nanometers and a very homogeneous, dense morphology. The grains exhibit smoothed cubic-like shapes and display essentially a disposition parallel to the substrate. This condition may be correlated with a preferential alignment along specific planes, as suggested by the XRD pattern analysis. Cross-sectional images have been also performed to assess the film thickness, resulting in an average thickness of 3.5 ± 0.1 µm (Fig. 6(c)).

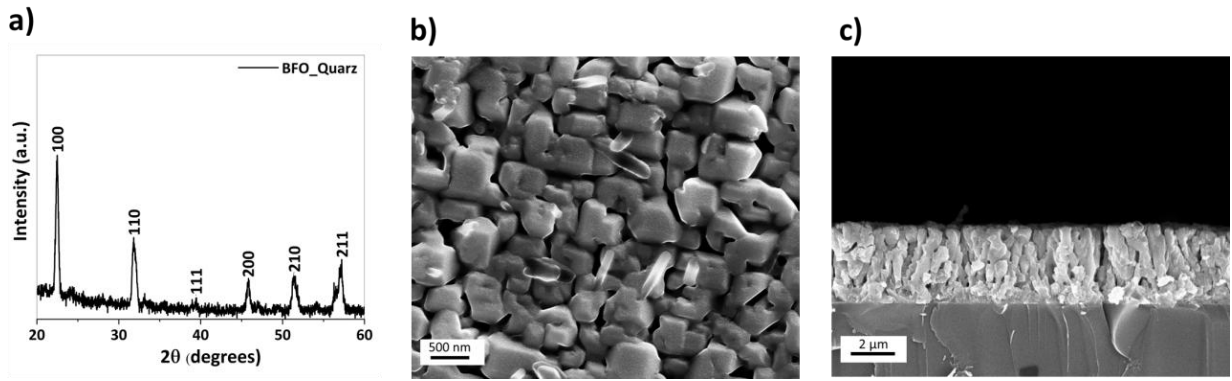

*Figure 6: (a) XRD pattern, (b) FE-SEM and (c) cross-sectional FE-SEM images of BFO film deposited on $SiO_2$ substrate.*

## 5. Permittivity and permeability of the $SiO_2$ substrate

In Fig. 7(a) the time dependent reference signal to calculate the complex transmission of $SiO_2$ substrate is presented as a black line. The red line represents instead the time dependent signal transmitted through the $SiO_2$ substrate, approximately 1 mm thick. Time dependent reflected signals are shown in Fig. 7(b). Black line is the reference signal acquired through an Au mirror whereas the red curve represents the signal reflected from the $SiO_2$ substrate. The transmission modulus $|\tilde{T}_s|$ and its phase (Arg) are shown in Fig. 7(c) as black curves, whereas the red one represents the modulus smoothed trend obtained through adjacent averaging (see Appendix I).

The modulus and the argument of the complex reflection function $|\tilde{R}_s|$ are reported as black curves in Fig. 7(d) whereas the red curve represents the modulus smoothed trend. Arguments of both $\tilde{T}_S^{(e)}$ and $\tilde{R}_S^{(e)}$ are not averaged because basically they do not present substantial FP oscillations. In Fig. 8(a) and (b) the analytical solutions provided by eqs. (5) and (6) representing the gross estimation of complex impedance and the refractive index respectively are reported. In Fig. 8(c) a typical contour plot of the error function to establish the best $\tilde{z}$ at a given frequency is shown. The high quality of the retrieval procedure is appreciable in Fig. 8(d), where the extracted parameters $(\tilde{n}_{s,R}, \tilde{z}_{s,R})$ and $(\tilde{n}_{s,T}, \tilde{z}_{s,T})$ are used to compare the theoretical reflection and transmission (modulus and argument) with the experimental counterparts.

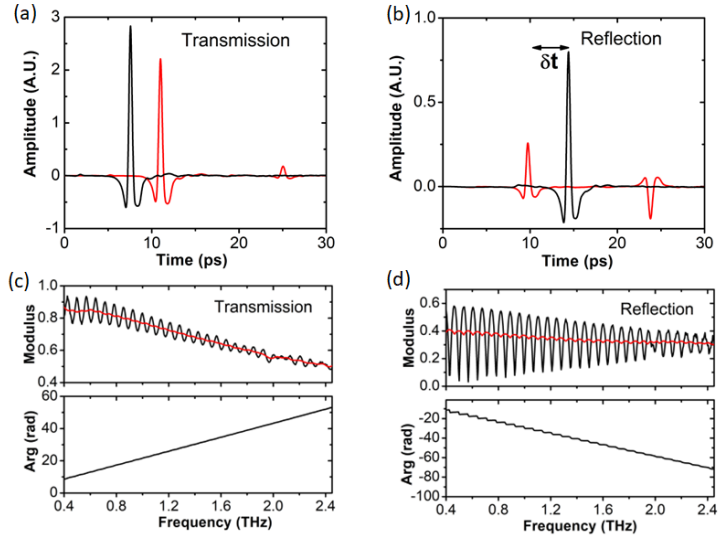

*Figure 7: (a) Time dependent signals transmitted through air (reference, black line) and the quartz substrate (red line). (b) Time dependent signals reflected by the Au film (reference, black line) and the quartz substrate (red line). (c) Modulus and argument of the transmission spectrum through the SiO$_2$ substrate (black lines) and its average expression (red line). (d) Modulus and argument of the reflection spectrum from the SiO$_2$ substrate (black lines) and its averaged values (red line).*

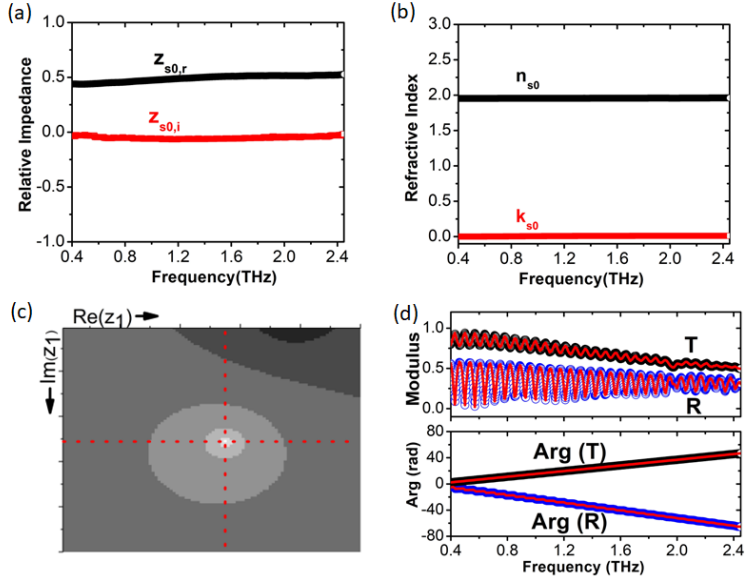

Figure 8: (a) The analytically retrieved complex impedance and (b) refractive index of the SiO$_2$ substrate. (c) Example of the error function contour plot showing the best $\tilde{z}$ highlighted by the cross of the two red dotted lines. (d) Comparison between the model (red curves) and the experimental transmission (black circles) and reflection (blue circles) behavior.

The output of the CTVT procedure applied to the SiO$_2$ substrate is reported in Figs. 9(a) and (b). Optimized impedance and refractive index differ by far less than one percent from the respective analytical solution. This behavior is owed to the fact that the FP contributions of a homogeneous optically thick slab do not contain additional information to improve the impedance and the refractive index with respect to the fundamental signal. This essentially implies the overlap of both $\tilde{\varepsilon}_T = \tilde{n}_T / \tilde{z}_T$ with $\tilde{\varepsilon}_R = \tilde{n}_R / \tilde{z}_R$ and $\tilde{\mu}_T = \tilde{n}_T \tilde{z}_T$ with $\tilde{\mu}_R = \tilde{n}_R \tilde{z}_R$, as shown in Fig. 9(c) and (d). The superposition between electrodynamics parameters, independently on the routine based on $\tilde{T}_s$ or $\tilde{R}_s$, suggests that the angle of misalignment ($\theta_m$) between the incident wave-vector direction and the normal to the sample surface is small enough to avoid any mismatch between the parameters retrieved through transmission and reflection measurements. The accuracy in the thickness value of the SiO$_2$ slab is extremely high, in the order of 0.5 %. This translates into an error bar unnoticeable on the scale of the panels (a) and (b) of Fig. (9). It is evident that the electrodynamic parameters of the SiO$_2$ slab fulfils the expected behavior of a full insulator material, showing $\varepsilon_{s,r} > \varepsilon_{s,i}$ and $\tilde{\mu} = 1$.

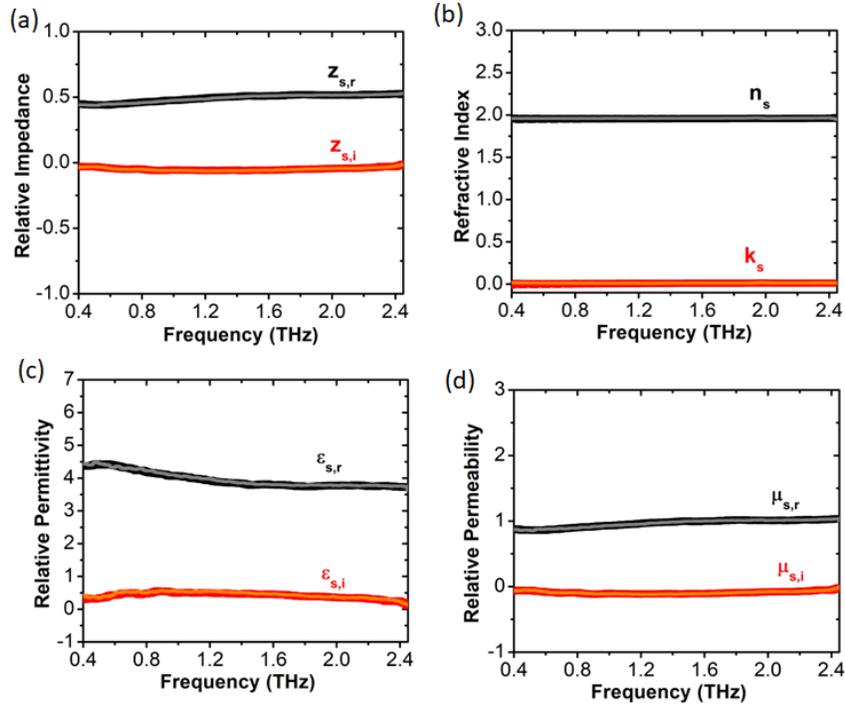

Figure 9: (a) The retrieved complex impedance and (b) refractive index of the $SiO_2$ substrate. Black and red curves refer to the outcomes obtained through the minimization of $Err(\tilde{T}_s)$, i.e. $\tilde{z}_{s,T}$ and $\tilde{n}_{s,T}$, whereas the grey and orange lines show the results achieved by the minimization of $Err(\tilde{R}_s)$ i.e. $\tilde{z}_{s,R}$ and $\tilde{n}_{s,R}$. (c) the complex permittivity and (d) permeability of the $SiO_2$ substrate. Black and red curves refer to $\tilde{\varepsilon}_T = \tilde{n}_T/\tilde{z}_T$, $\tilde{\mu}_T = \tilde{n}_T\tilde{z}_T$ whereas grey and orange curves describe $\tilde{\varepsilon}_R = \tilde{n}_R/\tilde{z}_R$, $\tilde{\mu}_R = \tilde{n}_R\tilde{z}_R$.

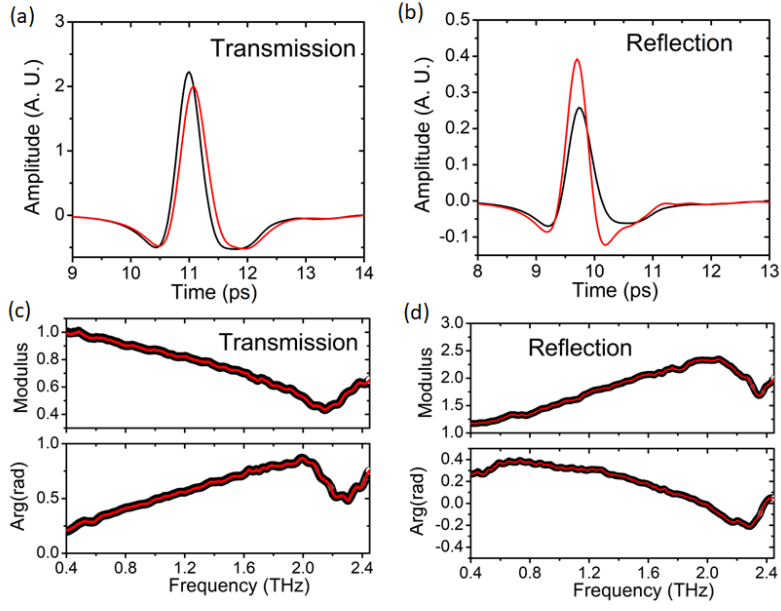

*Figure 10: (a) Time dependent transmitted and (b) reflected signals for the quartz substrate (black curve) and the BFO film (red curve). (c) Modulus and (d) argument of the transmission and reflection functions for the BFO film. Red lines represent the theoretical expectations using as parameters the best $\tilde{z}_f$ and $\tilde{n}_f$ retrieved via the CTVT method.*

## 6. Permittivity and permeability of BFO

Time dependent signals acquired through the transmission and the reflection from a BFO film $t = 3.5 \pm 0.1\ \mu m$ thick representative of the sample batch are reported in Figs. 10(a) and (b) respectively. Black curves represent the signals relative to the $SiO_2$ substrate whereas red curves indicate the measurements acquired in presence of the film. Averaged complex transmission ($\tilde{T}_f$) and reflection ($\tilde{R}_f$) are reported in terms of modulus in Fig. 10(c) and argument in Fig. 10(d). The high accuracy of the retrieval procedure can be appreciated observing the red lines shown in the figures, representing the theoretical expectations obtained for the $\tilde{T}_f$ and $\tilde{R}_f$ complex functions by inserting the optimized pairs ($\tilde{n}_{f,T}$, $\tilde{z}_{f,T}$) and ($\tilde{n}_{f,R}$, $\tilde{z}_{f,R}$) respectively, as explained below.

By following the same scheme reported in Fig. 2, eqs. (11) and (12) have been used to achieve the analytical expressions of impedance and refractive index of the film. Then, employing the CTVT and using eqs. (9) and (10), the best values of $\tilde{z}_f$ and $\tilde{n}_f$ are extracted and shown, along with their uncertainties in Fig. 11(a) and (b) respectively. Colors used for the lines are the same as for the $SiO_2$ substrate in Fig. 9, namely black and red curves are achieved through the minimization of $Err\{\tilde{T}_f\}$ whereas grey and orange curves are achieved through the minimization of $Err\{\tilde{R}_f\}$.

Differently from the case of the bare $SiO_2$ substrate, the impedance of BFO shows some dependence on the reflection or transmission minimization procedure. $\tilde{z}_{f,R}$ is slightly larger with respect to $\tilde{z}_{f,T}$ for frequency lower than $f_0 \sim 1.2\ THz$ (fig. 11(a)). Above the latter frequency the two impedances merge, implying that the mismatch for $f < f_0$ cannot be attributed to some misalignments. As a matter of fact, the refractive indexes $\tilde{n}_{f,R}$ and $\tilde{n}_{f,T}$, which are more sensitive to misalignments being included in the argument of the propagation factors, match within the error (fig. 11(b)). Therefore, ruling out any misalignment of the impinging beam with respect to the sample surface, we argue that the partial mismatch between $\tilde{z}_{f,R}$ and $\tilde{z}_{f,T}$ should be attributed to the gross approximation used to calculate the analytical solution $\tilde{z}_{f0}$. The latter quantity has been achieved by neglecting in the model the reflected beam originating at the film-substrate interface, namely the second addend in eq. (12), so yielding an underestimated film impedance. This problem is strictly related to the film thickness, since the thinner the film the higher the information on the second interface provided by the reflected beam. We have verified that the value of $\tilde{z}_f$ providing the best minimization of both $Err\{\tilde{T}_f\}$ and $Err\{\tilde{R}_f\}$ is confident with the average expression given by $\tilde{z}_{f,R}$ and $\tilde{z}_{f,T}$. However, as shown in Fig. 11(c) and (d), the retrieved $(\tilde{\varepsilon}_T, \tilde{\varepsilon}_R)$ and $(\tilde{\mu}_T, \tilde{\mu}_R)$ lie relatively close one to each other, so defining a precise region in which the BFO electrodynamics is clearly defined and arguable.

## 4. Discussion

As shown in Fig. 11(a), in the frequency range of observation the real part of the BFO impedance $z_{f,r}$ decreases from about 0.55 to 0.20 at $f_z \sim 2.2$ THz, where a minimum appears. From the minimum, $z_{f,r}$ increases up to 0.4 at around 2.35 THz. The corresponding imaginary part $z_{f,i}$ is negative and presents a specular behavior with respect to $z_{f,r}$.

The sign of $z_{f,i}$ can be interpreted in terms of a dominant "capacitive" behavior of the material. Nevertheless, the model of the optical impedance cannot neglect the inductive contribution as well, somehow related to the free electron contribution. In a previous work [48], we have shown that a reasonable description of the impedance $\tilde{Z}$ of a conducting material can be obtained inserting in parallel to the Drude branch, given by the series impedance of a resistance $R$ and an inductive reactance $L$, $\tilde{z}_D = R + i\omega L$, a capacitive contribution $C$, so that $\tilde{Z} = \tilde{z}_D // \tilde{z}_C$, where $\tilde{z}_C = -i/\omega C$. As long as $1/2\pi(LC)^{0.5}$ and $R/L$ are larger than the maximum frequency band, the negative sign of $Im(\tilde{Z})$ grows accordingly with the ratio $C/L$. Since $C$ is directly proportional to $\varepsilon_r$, in the present study negative values of the imaginary parts of both the substrate and the film fulfilling $|z_{f,i}| > |z_{s,i}|$ are justified by the fact that $\varepsilon_{f,r} > \varepsilon_{s,r}$. The latter inequality can be qualitatively understood by the presence of spontaneous electronic polarization in BFO that increases the real part of permittivity with respect to the substrate, given by the polarization induced by the presence of the impinging THz electric field only. BFO refractive index shows a real part $n_f$ varying in the interval $9 \div 10$ up to about 2.0 THz. In the same frequency band, the extinction coefficient $k_f$ varies in the range $1 \div 3$. For larger frequencies a resonating mechanism affects both real and imaginary parts of $\tilde{n}_f$ [24] and will be better discussed in the following paragraphs.

The electrical response of BFO to the THz field is provided by the permittivity reported in Fig. 11(c). Both real and imaginary part of permittivity $\varepsilon_{f,i}$ shows a clear resonance at about 2 THz in agreement with most of the papers appeared in literature, where the coupling of radiation with the BFO lattice is extensively discussed [19] [25]. However, $\varepsilon_{f,r}$ and $\varepsilon_{f,i}$ differ in value from previous findings [19] [24] [25], where permeability is usually set to $\tilde{\mu} \sim 1$. The magnetic behavior of BFO at THz frequencies is reported in Fig. 11(d). The real part of permeability $\mu_{f,r}$ is positive and from the value at about $\mu_{f,r} \sim 6$ decreases down to about $\mu_{f,r} \sim 1.3$ at 2.3 THz, where a resonance is detected. The imaginary part $\mu_{f,i}$ has a specular trend. It remains negative up to the resonance, whose maximum falls in the slightly positive range. The paramagnetic behavior of BFO in the THz region is consistent with the positive, though very small, magnetization measured through dc magnetometry [1]. We speculate that the robust enhancement of $\mu_{f,r}$ in this frequency band is produced by the strong magnetic coupling observed below 1 THz, as extensively discussed in refs. [24] and [26]. Certainly, in our measurements we observe magnetoelectric effects [49] [50], where magnetic and electric responses are mutually coupled so entailing the conversion of **H**-induced eddy currents in **E**-induced ones [51]. Multiferroics as BFO present dc magnetoelectric effects [6] but THz coupling with cycloid spins is backed for frequency lower than 1.5 THz where spin-wave excitations have been detected [26]. Thus, the magnetoelectric effect should occur to shift down $\mu_i$ towards negative values even in the band $f > 1.5\ THz$, where no magnetic coupling is expected. Yet, the signature of magnetoelectric effect should be visible in the opposite trends of $\varepsilon_r$ and $\mu_r$ [52]. By analyzing the spectrum before the phononic resonance ($f < 2\ THz$), $\varepsilon_r$ tends to increase despite $\mu_r$. The opposite trends of $\varepsilon_r$ and $\mu_r$ persists in correspondence of the phononic resonance ($f \geq 2\ THz$), albeit the frequency position of the resonance in $\tilde{\mu}$ shows a blueshift of about 0.25 THz. This is better shown in Fig. 12(a) and (b), where an expanded view around the phononic resonance of the averaged values of $\tilde{\varepsilon}$ and $\tilde{\mu}$ are plotted.

The frequency shift between the phononic resonances might be empirically interpreted as the way the lattice vibration induces different electric and magnetic resonances. The former would be related to the coupling with localized electrical dipoles ($\varepsilon_r$) and free charge motion ($\varepsilon_i$) whereas the latter would describe the coupling with the localized magnetic moments ($\mu_r$) and development of eddy currents ($\mu_i$). However, we argue that the frequency shift is an intrinsic consequence of the physical origin of transmission and reflection. As a matter of fact, the minima of $T$ and $R$ (Fig. 10(c)) are exactly positioned in correspondence of the resonances observed in $\tilde{\varepsilon}$ and $\tilde{\mu}$. Specifically, reflection is given by the sum of two back propagating waves [eq. (12)] which are phase shifted by the propagation factor relative to the back-and-forth propagation of the reflected signal inside the film. Such phase-lag (also mentioned in [53] [54]) between the two contributions in $\tilde{R}$ introduces a frequency shift with respect to the $\tilde{T}$ resonance.

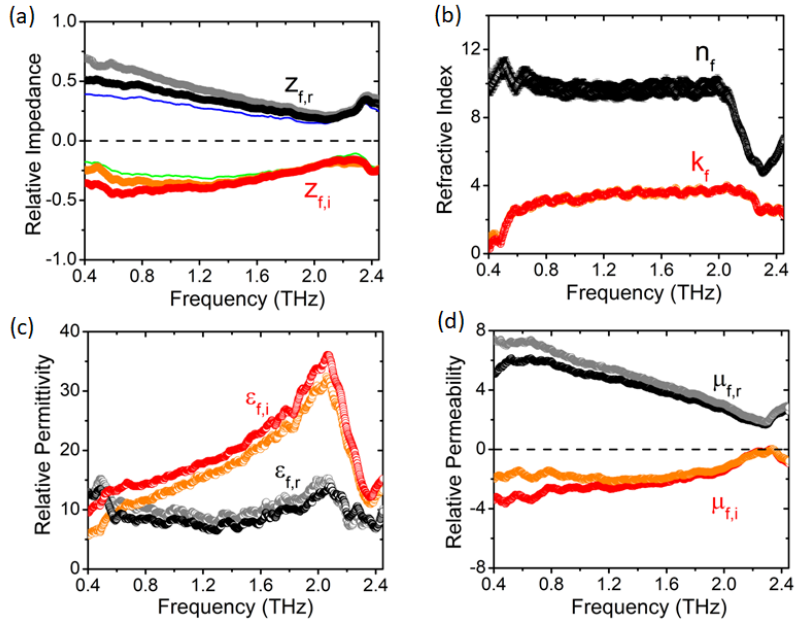

*Figure 11: Retrieved electrodynamic parameters obtained for the BFO film. (a) Relative impedance. The blue and green curves represent the real and imaginary parts of $\tilde{z}_{f0}$, respectively. (b) Complex refractive index. (c) Complex relative permittivity. (d) Complex relative permeability. In all panels black and red (grey and orange) curves refer to the results achieved by the minimization process of $Err\{\tilde{T}_f\}$ ($Err\{\tilde{R}_f\}$). Only in panel (b) the two procedures practically overlap.*

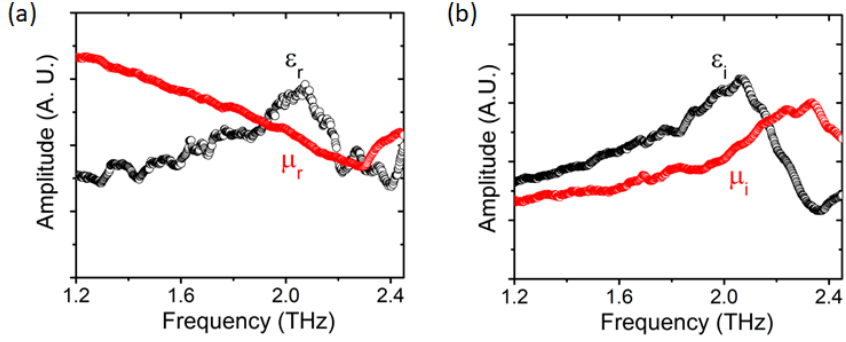

*Figure 12: (a) Real and (b) imaginary parts of permittivity (black open points) and permeability (red open points) of a BFO film. The permeability has been shifted to better show the comparison.*

A negative value for imaginary term in permeability is a relevant issue regarding the ways the system can dissipate the impinging EM waves. In general, the imaginary parts of permittivity and permeability are expected to be positive because they describe the energy provided by the EM waves to excite electrons whose relaxation produces heat [50]. Nevertheless, negative $\mu_i$ has been previously observed in different systems ( [51] [52] [53] [55] [54] and references therein). The experimental evidence that $\mu_i < 0$ can be accounted for using one or both of the following arguments:

i.  Magnetic energy is radiated out of the sample thus disabling any heating mechanism [51] [53];
ii. In transmission and reflection measurements the presence of FP resonances introduces artifacts causing $\mu_{f,i} < 0$ [55];

The argument *i.*) can be used when the sample is composed of particles with a transversal dimension much lower than the impinging wavelength. $E$ and $H$ currents cannot expand within the particle size and so radiate out. This mechanism can even induce $\varepsilon_i < 0$ [51]. In our experiments, $E$ impinges over a large film area unlike $H$, that fluxes across a very thin area $S \times t$, where $S$ is the spot size. Thus, *i.*) makes sense since the inequality $\lambda \gg t$ is verified for the whole frequency band. The second argument is usually set out in the microwave band when the sample is in the order of a few mm as in the case discussed in [55]. In this work, the FP modulation in the film is negligible, therefore we believe that this hypothesis should be discarded.

Since the phenomenon finds a reasonable explanation by the first argument, it is then necessary to figure out if $\mu_{f,i} < 0$ is a condition expected by the sign of real and imaginary parts of $\tilde{z}_f$ and $\tilde{n}_f$, all positive but $z_{f,i}$. Yet, we have previously discussed that for a plane film system it is also expected that $\varepsilon_{f,i} > 0$. Indeed, $\mu_{f,i} < 0$ is compatible with a corresponding positive value for the imaginary part of permittivity, as it turns out from the dependence of $\varepsilon_{f,i}$ and $\mu_{f,i}$ on $\tilde{n}$ and $\tilde{z}$:

$$\varepsilon_{f,i} = k_f z_{f,r} - n_f z_{f,i} \qquad (17)$$

$$\mu_{f,i} = k_f z_{f,r} + n_f z_{f,i}. \qquad (18)$$

From the positive sign of eq. (17) one obtains

$$z_{f,i} < (k_f/n_f) z_{f,r} \qquad (19)$$

that is always fulfilled by a capacitive impedance because the right member is definitely positive. From eq. (18), being $\mu_{f,i} < 0$ the imaginary part of the impedance must satisfy

$$z_{f,i} < -(k_f/n_f)z_{f,r} \qquad (20)$$

that is compatible with the capacitive feature ($z_{f,i} < 0$) of the film impedance.

Hence, if the imaginary part of the impedance has a capacitive behavior, the condition $\mu_{f,i} < 0$ is compatible with the required condition $\varepsilon_{f,i} > 0$. Since our results comply with $Im\{\tilde{\varepsilon}\} + Im\{\tilde{\mu}\} > 0$, the second law of thermodynamics is not violated [51] [56] and the results summarized in Fig. 11 are physically meaningful.

## 8. Conclusions

BFO samples prepared using a CVD technique are characterized at room temperature via TDS-THz spectroscopy in both normal transmission and reflection configurations. An accurate retrieval procedure, based on modelling the complex functions $\tilde{T}$ and $\tilde{R}$ and matching the experimental frequency data with the expected theoretical behavior, is introduced to achieve both the permittivity and permeability of the film. Values are extracted through independent routines providing $\tilde{\varepsilon}_T$, $\tilde{\varepsilon}_R$, $\tilde{\mu}_T$, $\tilde{\mu}_R$, which identify a precise trend (with relative uncertainty) for the electric and magnetic responses. Both the real part of $\tilde{\varepsilon}$ and $\tilde{\mu}$ show a substantial monotonic frequency dependence up to about 2 THz, above which a phononic resonance is observed. The imaginary part of permeability instead shows negative values in the frequency range of investigation. Above 2 THz, the EM waves couple with the lattice and consequently a phononic resonance is observed in both the electric and magnetic responses. A frequency shift of about 0.25 THz between the $\tilde{\varepsilon}$ and $\tilde{\mu}$ observed resonances is explained by the phase-lag of waves bouncing back and forth off the two interfaces of the film. The magnetoelectric effect appears to affect the trend of real parts of permittivity and permeability but the negative sign of the imaginary part of permeability is more likely related to the magnetic energy scattered from the film. The negative sign of $\mu_{f,i}$ does not violate the second law of thermodynamics provided that the sum $\varepsilon_{f,i} + \mu_{f,i}$ is fully positive.

## Appendix I

We show that the use of the adjacent averaging (AA) method is very effective in removing the sinusoidal contribution from the transmitted or reflected signal because its process cannot be biased by any choice of the user.

Time domain THz spectroscopy of optically thick samples always suffers for the visible presence of Fabry-Perot (FP) oscillations generated by the multiple reflections of the signal in the sample itself. FP effects in the frequency dependent transmitted signal ($\tilde{T}$) can introduce relevant distortions which can lower the accuracy of the retrieved EM parameters. Some remarkable approach to efficiently deal with this problem may be found in [57] [58] [59] which enable to extract the FP contribution from the complex transmission. Withayachumnankul *et al.* in [57] introduce a methodology consisting into fitting the logarithm of $\tilde{T}$ through a polinomial function that enables to evaluate the FP contribution. In [58] the error function defined by the difference between the experimental and theoretical transfer function is transformed in a series of differential equations enabling to find the EM parameters deprived of the FP contribution, whereas in [59] the FP removal is achieved through a methodology running in time domain. In all these studies however the transfer function is exclusively dependent on the refractive index $\tilde{n}$, namely it can be applied only to materials where the assumption $\tilde{\mu} \approx 1$ is acceptable.

Our approach to remove FP oscillation is simple and enables to achieve accurate expressions of the EM parameters. The goal is to obtain transfer functions $\tilde{T}$ and $\tilde{R}$ in which the sinusoidal contribution is accurately removed trying to avoid any distortion in the monotonic background. Then the CTVT will provide the accurate expressions of the EM parameters, namely $\tilde{z}$ and $\tilde{n}$.

The approach we propose lays on the assumption that whatever is the transfer function $F$, it can be expressed at a given frequency $\omega_i$ as

$$F(\omega_i) = f(\omega_i) + s(\omega_i + \phi) \tag{1a}$$

where $f(\omega_i)$ is the discrete background function, $s(\omega_i)$ is the sinusoidal contribution induced by the FP effect, with $\phi$ the starting phase of $s$. It has to be noted that $f(\omega_i)$ is not necessarily a monotonous function, provided that the changes in derivative are not as fast as the ones showing up in the FP oscillations.

A standard AA technique applied to $F(\omega_i)$ yields:

$$< F(\omega_i) >_N = \sum_{j=\omega_i-N/2}^{\omega_i+N/2} F(\omega_j) \approx f(\omega_i). \tag{2a}$$

The approximate nature of eq. (2a) has three main reasons:

*(i)* Discretization. The inherent character of the functions we are dealing with, unavoidably introduces residual contributions mostly depending on the missing proportionality between the sampling period and the period of the $s$ function.

*(ii)* Starting phase $\phi$. This will induce the $s$ contribution to display an uncomplete oscillation that is not going to be averaged to zero.

*(iii)* Band edges. In agreement with eq. (2a) the first and the last $N/2$ points cannot be processed through AA.

The error discussed in *(i)* decreases if the sampling period does, whereas the error in *(ii)* falls in the *(iii)* case because $N$ is of the order of the $s$ period and $\phi$ is of the order of $N/4$ points, i.e. ¼ of a period of a sinusoid.

The main scope of this Appendix is to figure out that exists the best $N$ related to the period of $s$, say $N_T$, providing the minimum residual from the average process. Without losing generality, let's suppose we want to process a sinusoid in the center of the operational frequency band so that we can neglect the error sources *(ii)* and *(iii)*. It is straightforward to realize that if we set

$$\sum_{i=-N_T/2}^{N_T/2} s(\omega_i) = min(s) \tag{3a}$$

then any other averaging interval, say $N = N_T \pm 2$, will provide

$$\sum_{i=-\frac{N_T}{2}\pm 1}^{\frac{N_T}{2}\pm 1} s(\omega_i) = min(s) + \sum_{i=\pm 1}^{\pm 1} s(\omega_i) = min(s) + r(2) \tag{4a}$$

where $r(2)$ is the residual error for 2 points owed to the mismatch between $N$ and $N_T$. The residual is a periodic function of $N - N_T$ that presents a maximum for $|N - N_T| \sim N_T/4$. Eqs. (3a) and (4a) summarize the main aspect of the issue on the AA process regarding the presence of a unique $N_T$ that minimizes the sinusoidal contribution.

In conclusion, the transmitted and reflected signal achieved from an experiment are defined with a specific number of points defining the period of the sinusoidal contribution and consequently the number of points adopted to perform the AA process are univocally determined.